\documentclass[twocolumn,superscriptaddress,notitlepage]{revtex4-2}

\usepackage{amsmath,amssymb,setspace}
\usepackage{verbatim}
\usepackage{physics}
\usepackage{graphicx}
\usepackage{newtxtext}
\usepackage{newtxmath}
\usepackage[utf8]{inputenc}
\usepackage[x11names]{xcolor}
\usepackage[unicode,colorlinks,citecolor=Blue3,linkcolor=Blue3,bookmarks=true]{hyperref}

\newcommand{\mean}[1]{\langle{#1}\rangle}
\newcommand{\svector}[2]{\begin{pmatrix}#1 \\ #2 \end{pmatrix}}
\newcommand{\smatrix}[4]{\begin{pmatrix}#1 & #2 \\ #3 & #4\end{pmatrix}}

\begin{document}

\title{Improving Kerr QND Measurement Sensitivity via Squeezed Light}

\author{Stepan Balybin}
\email{sn.balybin@physics.msu.ru}
\affiliation{Faculty of Physics, M.V.Lomonosov Moscow State University, Leninskie Gory 1, Moscow 119991, Russia}
\affiliation{Russian Quantum Center, Skolkovo IC, Bolshoy Bulvar 30, bld.\ 1, Moscow, 121205, Russia}

\author{Dariya Salykina}%
\affiliation{Faculty of Physics, M.V.Lomonosov Moscow State University, Leninskie Gory 1, Moscow 119991, Russia}
\affiliation{Russian Quantum Center, Skolkovo IC, Bolshoy Bulvar 30, bld.\ 1, Moscow, 121205, Russia}

\author{Farid Ya.\ Khalili}
\email{farit.khalili@gmail.com}
\affiliation{Russian Quantum Center, Skolkovo IC, Bolshoy Bulvar 30, bld.\ 1, Moscow, 121205, Russia}

\begin{abstract}

In Ref.\,\cite{21a1BaMaKhStIlSaLeBi}, the scheme of quantum non-demolition measurement of optical quanta that uses a resonantly enhanced Kerr nonlinearity in optical microresonators was analyzed theoretically. It was shown that using the modern high-$Q$ microresonators, it is possible to achieve the sensitivity several times better than the standard quantum limit. Here we propose and analyze in detail a significantly improved version of that scheme. We show, that by using a squeezed quantum state of the probe beam and the anti-squeezing (parametric amplification) of this beam at the output of the microresonator, it is possible to reduce the measurement imprecision by about one order of magnitude. The resulting sensitivity allows to generate and verify multi-photon non-Gaussian quantum states of light, making the scheme considered here interesting for the quantum information processing tasks.

\end{abstract}

\maketitle

\section{Introduction}

An ideal quantum measurement described by von Neumann’s reduction postulate \cite{Neumann_e} does not perturb the measured observable $N$. The sufficient condition for implementation of such a measurement is the commutativity of the operator $\hat{N}$ with the Hamiltonian $\hat{H}$ of the combined system ``measured object+meter'' \cite{Bohm1951e, 92BookBrKh}:
\begin{equation}\label{QND_cond_1}
  [\hat{N},\hat{H}] = 0 \,,
\end{equation}
where
\begin{equation}
  \hat{H} = \hat{H}_S + \hat{H}_A + \hat{H}_I \,,
\end{equation}
$\hat{H}_S$, $\hat{H}_A$ are, respectively, the Hamiltonians of the object and the meter, and $\hat{H}_I$ is the interaction Hamiltonian. In the article \cite{Thorne1978}, the term ``quantum non-demolition (QND) measurement'' was coined for this type of measurement.

In many cases, a sequence of measurement of a variable $N(t)$ is required, instead of a single measurement. The typical example is detection of external force acting on the object. In this case the value of $\hat{N}$ have to be preserved between the measurements, which leads to another (also sufficient) commutativity condition:
\begin{equation}\label{QND_cond_2}
  [\hat{N},\hat{H}_S] = 0 \,,
\end{equation}
The observables which satisfy both conditions \eqref{QND_cond_1} and \eqref{QND_cond_2} are known as QND observables.

It follows from Eqs.\,\eqref{QND_cond_1},\,\eqref{QND_cond_2} that the interaction Hamiltonian have to commute with the measured observable:
\begin{equation}\label{QND_cond_I}
  [\hat{N},\hat{H}_I] = 0 \,,
\end{equation}
In the particular case of the electromagnetic energy or number of quanta measurement, this means that nonlinear interaction of the electromagnetic field with the meter has to be used. A semi-gedanken example of such a measurement based on the radiation pressure effect was considered in Ref.\,\cite{77a1eBrKhVo}.

Later a more practical scheme based on the cubic (Kerr) optical non-linearity \cite{Milburn_PRA_28_2065_1983} was proposed. In this scheme, the signal optical mode interacts with another (probe) one by means of the optical Kerr nonlinearity. As a result, the phase of the probe mode $\varphi_p$ is changed depending on the photon number $N_s$ in the signal one (the cross phase modulation, XPM). The subsequent  interferometric measurement of this phase allows to retrieve the value of $N_s$ with the precision depending on the initial uncertainty of $\varphi_p$, see details in Sec.\,IV of Ref.\,\cite{21a1BaMaKhStIlSaLeBi}. In the ideal lossless case, the photon numbers in both modes are preserved. However, due to the XPM effect, the phase of the signal mode is perturbed proportionally to the probe mode energy uncertainty, fulfilling thus the Heisenberg uncertainty relation.

The natural sensitivity threshold for the QND measurement of the number of quanta is the standard quantum limit (SQL)
\begin{equation}\label{SQL}
  \Delta N_{\rm SQL} = \sqrt{N} \,,
\end{equation}
where $N$ is the mean number of the measured quanta. It corresponds to the best possible sensitivity of a linear non-absorbing meter \cite{92BookBrKh}, for example a phase-preserving linear amplifier \cite{Heffner_ProcIRE_50_1604_1962}. Starting from the initial work \cite{Levenson_PRL_57_2473_1986}, many proof-of-principle experiments based on the XPM  idea were done, see the review articles \cite{Roch_APB_55_291_1992, 96a1BrKh, Grangier_Nature_396_537_1998}. The sensitivity  exceeding the SQL was demonstrated in these experiments, but the ultimate goal of the single-photon accuracy was not reached due to the insufficient values of the optical nonlinearity in highly transparent optical media.

The recent advantages in fabrication of high-$Q$ monolithic and integrated microresonators \cite{Strekalov_JOptics_18_123002_2016}, which combine very low optical losses with high concentration of the optical energy promises the way to overcome this problem. This possibility was analyzed  in detail in Ref.\,\cite{21a1BaMaKhStIlSaLeBi}. It was shown that the sensitivity of the Kerr nonlinearity based QND measurement schemes is limited by the interplay of two undesirable effects: the optical losses and the self phase modulation (SPM) of the probe mode, which perturbs the probe mode phase proportionally to the energy uncertainty of this mode.

It was shown in Ref.\,\cite{Drummond_PRL_73_2837_1994}, that in the ideal lossless case, the SPM effect can be compensated using the measurement of the optimal quadrature of the output probe field instead of the phase one. However, in presence of the optical losses, only partial compensation is possible, limiting the sensitivity by the value
\begin{equation}\label{dNs_prev}
  (\Delta N_s)^2 = \frac{1}{\Gamma_X^2}
    \biggl[\frac{1}{4\eta N_p} + (1-\eta)N_p\Gamma_S^2\biggr] ,
\end{equation}
see Eq.\,(32) of \cite{21a1BaMaKhStIlSaLeBi}. Here $\Delta N_s$ is the measurement error, $N_p$ is the photon number in the probe mode, $\Gamma_X$, $\Gamma_S$ are the dimensionless factors of, respectively, the cross and the self phase modulation, see Eqs.\,\eqref{g2G} and \eqref{chi2G}, and $\eta$ is the quantum efficiency of the measurement channel. The second term here stems from the SPM. Due to this term, the optimal value of $N_p$ exists, which provides the minimum of $\Delta N_s$:
\begin{equation}\label{dNs_min_prev}
  (\Delta N_s)^2 = \frac{\Gamma_S}{\Gamma_X^2}\epsilon \,,
\end{equation}
where
\begin{equation}
  \epsilon = \sqrt{\frac{1-\eta}{\eta}}
\end{equation}
is the normalized loss factor. It was shown in Ref.\,\cite{21a1BaMaKhStIlSaLeBi}, that using the best microresonators available now, the sensitivity $\Delta N_s\sim10^2\text{-}10^3$ could be achieved.

Eqs.\,\eqref{dNs_prev}, \eqref{dNs_min_prev} imply that the probe mode is prepared in the coherent quantum state. At the same time, in was shown by C.Caves in the work \cite{Caves1981}, that the sensitivity of optical interferometric measurements can be improved without increasing the optical power by using the non-classical squeezed states of light. Currently, this method is routinely used in the laser gravitational-wave detectors \cite{Tse_PRL_123_231107_2019_short, Acernese_PRL_123_231108_2019_short}; see also the review \cite{22a1SaKh}.

In the same work \cite{Caves1981}, C.\,Caves proposed also to use an additional degenerate optical parametric amplifier (anti-squeezer) at the output of the  interferometer to reduce impact of the losses in the optical elements located after this amplifier, including the photodetector(s) inefficiency. Note that usually, it is the output losses constitute the major part of the total losses budget in the optical interferometers. Recently, this method was demonstrated experimentally \cite{20a1FrAgKhCh}.

In the current work we show that sensitivity of the QND measurement scheme, considered in Ref.\,\cite{21a1BaMaKhStIlSaLeBi}, can be radically improved using these techniques. This paper is organized as follows. In Sec.\,\ref{sec:a2b}, we derive the linearized input/output relations for the microresonator. In Sec.\ref{sec:dN}, we calculate the measurement error, taking into account the SPM effect and the losses in the probe mode.  In Sec.\,\ref{sec:estimates}, we estimate the sensitivity, which can be achieved using the best available microresonators. In Sec.\,\ref{sec:mu} we consider the effect of optical losses in the signal mode both in the measurement and in the quantum state preparation scenarios. In Sec.\,\ref{sec:qc}, we discuss possible applications of the considered scheme to quantum information processing. We summarize the results of this paper in Sec.\,\ref{sec:conclusion}.

\section{Evolution of the optical fields in the microresonator}\label{sec:a2b}

\begin{figure}[t]
  \includegraphics[width=0.4\textwidth]{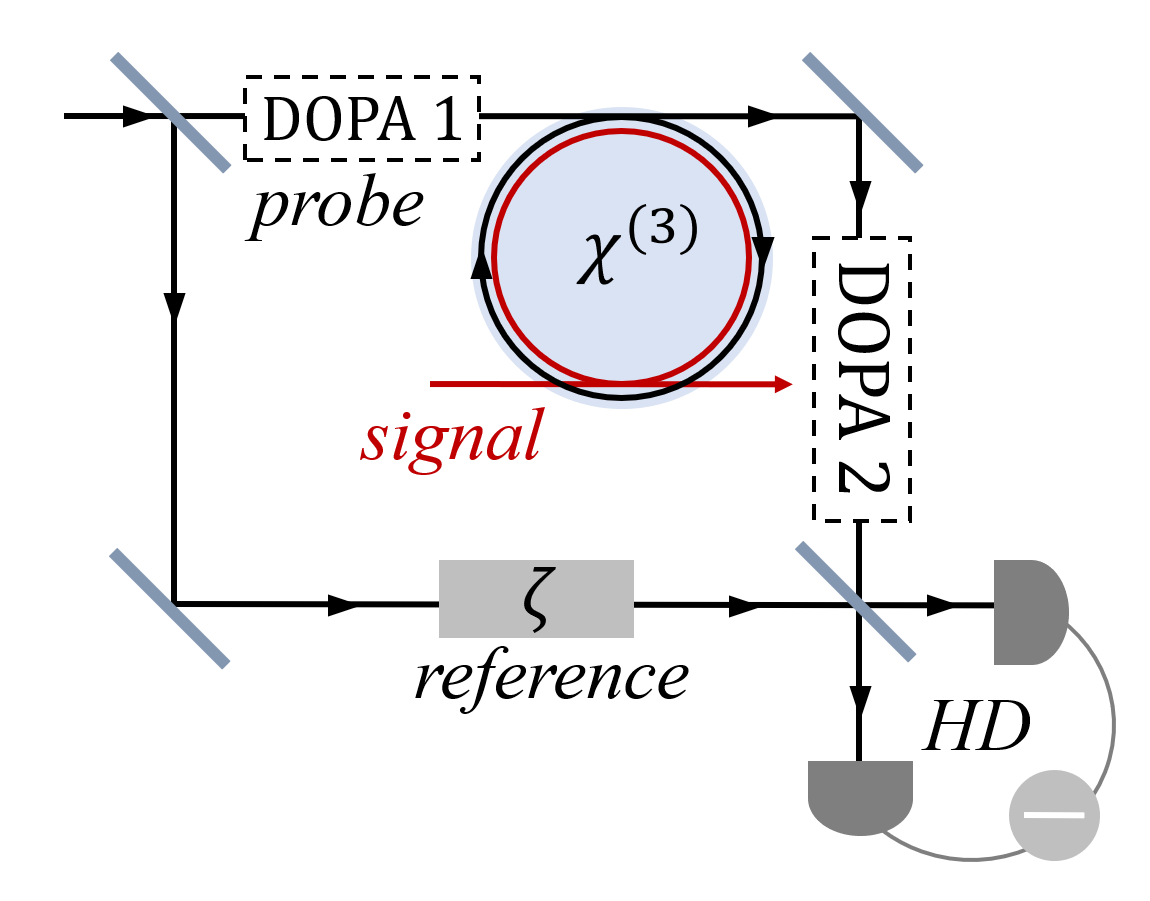}
  \caption{The scheme of the QND measurement of the photon number in signal mode via XPM effect. DOPA: degenerate optical parametric amplifier, HD: homodyne detector.}\label{fig:scheme}
\end{figure}

The measurement scheme is shown in Fig.\,\ref{fig:scheme}. Here the strong coherent beam is split by the unbalanced beamsplitter with the transmissivity $T\ll1$ into the probe and the reference beams. The probe beam is squeezed by the degenerate optical parametric amplifier DOPA\,1 and then injected into the microresonator. There it interacts with the signal beam and then passes through the second optical parametric amplifier DOPA\,2 and finally is recombined with the reference beam in the homodyne detector.

Taking into account that the main goal of this paper is to show the advantages provided by the squeezing, and in order to simplify the calculations, we, similar to the paper \cite{21a1BaMaKhStIlSaLeBi}, do not take into account explicitly the optical losses in the microresonator, but assume instead that
\begin{equation}\label{Q_load}
  \frac{Q_{\rm load}}{Q_{\rm intr}} \approx \frac{\tau}{\tau^*} \ll \epsilon^2 \,,
\end{equation}
where $Q_{\rm load}$ is the loaded quality factor, $Q_{\rm intr}$ is the intinsic one, $\tau$ is the interaction time, and $\tau^*$ is the relaxation time of the microresonator.

In this case, the interaction of the signal and probe modes in the non-linear microresonator can be described by the following Hamiltonian  \cite{21a1BaMaKhStIlSaLeBi}:
\begin{equation}
  \hat{H} = \sum_{x=p, s}\left[
      \hbar\omega_x\hat{N}_x - \frac{\hbar \gamma_S}{2}\hat{N}_x(\hat{N}_x-1)
    \right]
  - \hbar\gamma_X\hat{N}_p\hat{N}_s \,,
\end{equation}
where
\begin{equation}
  \hat{N}_{p,s} = \hat{a}_{p,s}^\dag\hat{a}_{p,s}
\end{equation}
are the photon number operators in the probe and signal modes, respectively, $\hat{a}_{p,s}$, $\hat{a}_{p,s}^\dag$ are the corresponding annihilation and creation operators for these modes, $\omega_{s,p}$ are their frequencies, $\gamma_S$, $\gamma_X$ --- coefficients of SPM and XPM interactions, respectively. Evidently, both $\hat{N}_p$ and $\hat{N}_s$ commute with this Hamiltonian and therefore are preserved.

The corresponding Heisenberg equations of motion for the operators $\hat{a}_s$, $\hat{a}_p$ are the following:
\begin{subequations}\label{eqs_sp}
  \begin{gather}
    \frac{d \hat{a}_p}{d t}
      = i\bigl(-\omega_p + \gamma_S\hat{N}_p + \gamma_X\hat{N}_s\bigr)\hat{a}_p\,,
      \label{eqs_p} \\
    \frac{d \hat{a}_s}{d t}
      = i\bigl(-\omega_s + \gamma_S\hat{N}_s + \gamma_X\hat{N}_p\bigr)\hat{a}_s\,.
  \end{gather}
\end{subequations}
The solution for the signal mode can be presented as follows:
\begin{equation}
  \hat{a}_s(t)
  = \exp\bigl[i\bigl(-\omega_s + \gamma_S\hat{N}_s + \gamma_X\hat{N}_p\bigr)t\bigr]
      \hat{a}_s \,,
\end{equation}
which clearly shows that its phase is perturbed by (i) the XPM effect, which is the inevitable consequence of the uncertainty relation, and by (ii) the parasitic SPM effect.

The similar equation can be written down also for the probe mode. However, here we take into account that in order to implement the high precision measurement, large mean number of quanta in the probe mode $N_p$ is required. This assumption allows to linearize the probe mode equation of motion and thus smoothly integrate it into the general framework of calculation of the optical interferometer.

We explicitly single out the part $\alpha_p$ in $\hat{a}_p$ that contains the  classical oscillations amplitude (that is, the average value of this operator):
\begin{equation}
  \hat{a}_p \to \alpha_p + \hat{a}_p \,.
\end{equation}
We suppose that classical amplitude is strong enough in comparison with quantum one:
\begin{equation}\label{big_N_p}
  N_p = |\alpha_p|^2 \gg 1 \,.
\end{equation}
Without limiting the generality, we assume that $\alpha_p$ is real.

In addition, we present the signal mode photon number as follows:
\begin{equation}
  \hat{N}_s + N_s + \delta\hat{N}_s \,,
\end{equation}
where $N_s$ is the mean value and $\delta\hat{N}_s$ is the a priori uncertainty, which we assume to be small:
\begin{equation}
  \delta N_s \ll N_p \,.
\end{equation}

With account for this, the classical part of Eq.\,\eqref{eqs_p} can be presented as follows:
\begin{equation}\label{eqs_p_0}
  \frac{d\alpha_p}{dt} =-i\omega_p^{\prime}\alpha_p \,,
\end{equation}
where
\begin{equation}
  \omega_p^{\prime} = \omega_p - \gamma_S\alpha_p^2 - \gamma_XN_s
\end{equation}
is the “dressed” eigenfrequency of the probe mode. Subtracting Eq.\,\eqref{eqs_p_0} from Eq.\,\eqref{eqs_p}, keeping only first order in $\hat{a}_p$, $\hat{a}_p^\dag$ terms, and using the rotated with the frequency $\omega_p'$ picture, we obtain the following linearized equation:
\begin{equation}\label{eqs_p_1}
  \frac{d \hat{a}_p}{d t} = i\bigl\{
      N_p\gamma_S[\hat{a}_p(t) + \hat{a}_p^\dag(t)] + \alpha_p\gamma_X\delta\hat{N_s}
    \bigr\} .
\end{equation}

Introduce now cosine and sine quadratures of the pump mode:
\begin{equation}
  \hat{a}_p^c  = \frac{\hat{a}_p + \hat{a}_p^\dag}{\sqrt{2}} \,,\quad
  \hat{a}_p^s  = \frac{\hat{a}_p - \hat{a}_p^\dag}{i\sqrt{2}} \,,
\end{equation}
In these notations, Eqs.\,\eqref{eqs_p_1} can be rewritten as follows:
\begin{subequations}
  \begin{gather}
    \frac{d \hat{a}_{p}^c(t)}{d t} = 0 \,, \\
    \frac{d \hat{a}_{p}^s(t)}{d t}
      = 2N_p\gamma_S\hat{a}_{p}^c(t) + \sqrt{2N_p}\gamma_X\delta\hat{N}_{s} \,.
  \end{gather}
\end{subequations}
Integrating these equations, we obtain the input/output relations for the probe mode. In the matrix notation, they can be presented as follows:
\begin{equation}\label{a2b}
  \svector{\hat{b}_p^c}{\hat{b}_p^s}
  \equiv \svector{\hat{a}_{p}^c(\tau)}{\hat{a}_{p}^s(\tau)}
  = \mathbb{F}\svector{\hat{a}_{p}^c}{\hat{a}_{p}^s}
    + \sqrt{2N_p}\Gamma_X\delta\hat{N}_s\svector{0}{1} \,,
\end{equation}
where
\begin{equation}
  \mathbb{F} = \smatrix{1}{0}{2N_p\Gamma_S}{1}
\end{equation}
is the SPM matrix, $\tau$ is the interaction time and the factors $\Gamma_{X,S}$ are equal to
\begin{equation}\label{g2G}
  \Gamma_{X,S}=\gamma_{X,S} \tau \,.
\end{equation}

\section{Squeezing and parametric amplification}\label{sec:dN}

We assume that the probe field is prepared in the squeezed coherent state. In this case, using again the matrix notations, its initial state can be presented as follows:
\begin{equation}\label{z2a}
  \svector{\hat{a}_p^c}{\hat{a}_p^s} = \svector{\sqrt{2\alpha_p}}{0}
    + \mathbb{S}(r,\theta)\svector{\hat{z}^c}{\hat{z}^s} ,
\end{equation}
where the quadratures $\hat{z}^{c,s}$ correspond to a ground state field and
\begin{equation}
  \mathbb{S}(r,\theta) = \smatrix{\cosh r + \sinh r\cos2\theta}{\sinh r\sin2\theta}
    {\sinh r\sin2\theta}{\cosh r - \sinh r\cos2\theta} .
\end{equation}
is the squeeze matrix.

Transformation of the probe mode at the output of the microresonator by the second DOPA (the anti-squeezer) can be described in the similar way:
\begin{equation}\label{b2c}
  \svector{\hat{c}_p^c}{\hat{c}_p^s}
    = \mathbb{S}(R,\phi)\svector{\hat{b}_c}{\hat{b}_s} ,
\end{equation}
where $\hat{c}_p$, $\hat{c}_s$ describe the probe field at the output of the DOPA\,2 and $R$, $\phi$ are the corresponding squeeze factor and squeeze angle.

The final step is the homodyne measurement of the quadrature
\begin{equation}\label{c_zeta}
  \hat{c}_p^\zeta = {\bf H}^{\sf T}(\zeta)\svector{\hat{c}_p^c}{\hat{c}_p^s} \,,
\end{equation}
defined by the homodyne angle $\zeta$, where
\begin{equation}
  {\bf H}(\zeta) = \svector{\cos\zeta}{\sin\zeta}
\end{equation}
is the homodyne vector.

We take into account the output losses, including the detection inefficiency, by using the model of the imaginary beamsplitter with the power transmissivity $\eta$, located before the detector (see \cite{Leonhardt_PRA_48_4598_1993}). It transforms the output signal as follows:
\begin{equation}\label{c2d}
  \hat{d}_p^\zeta=\sqrt{\eta}\,\hat{c}_p^\zeta + \sqrt{1-\eta}\,\hat{y} \,,
\end{equation}
where $\hat{y}$ it the corresponding quadrature amplitude of a vacuum field.

Combining Eqs.\,\eqref{a2b}, \eqref{b2c}, \eqref{c2d} and separating the signal term proportional to $\delta\hat{N}_s$ from other (noise) ones, we obtain the following equation for the output signal:
\begin{equation}
  \hat{d}_p^\zeta = \hat{d}_0^\zeta + G\,\delta\hat{N}_s \,,
\end{equation}
where
\begin{equation}\label{d0zeta}
  \hat{d}_0^\zeta
  = \sqrt{\eta}\,{\bf H}^{\sf T}(\zeta)\mathbb{S}(R,\phi)\mathbb{F}
      \svector{\hat{a}_c}{\hat{a}_s}
    + \sqrt{1-\eta}\,\hat{y}
\end{equation}
is the part of $\hat{d}_p^\zeta$ which does not depend on $\delta\hat{N}_s$ and
\begin{equation}\label{G}
  G = \sqrt{2\eta N_p}\Gamma_X{\bf H}^{\sf T}(\zeta)\mathbb{S}(R,\phi)\svector{0}{1}
\end{equation}
is the gain factor.

Therefore, the measurement error for the signal mode photon number is equal to
\begin{equation}\label{DNs}
  (\Delta N_s)^2 = \frac{(\Delta\hat{d}_{0}^\zeta)^2}{G^2} \,,
\end{equation}
where $(\Delta\hat{d}_{0}^\zeta)^2$ is the variance of $\hat{d}_0$. It depends on three parameters: the squeeze angles $\theta$, $\phi$, and the homodyne angle $\zeta$, which have to be optimized. This optimization is done in App.\,\ref{app:angles}, giving the following result:
\begin{equation}\label{result}
  (\Delta N_s)_{\rm meas}^2 = \frac{1}{\Gamma_X^2}\biggl(
    \frac{e^{-2r} + \epsilon^2e^{-2R}}{4N_p}
    + \frac{N_p\Gamma_S^2\epsilon^2}{e^{2R} + \epsilon^2e^{2r}}\biggr) .
\end{equation}

Comparison of Eqs.\,\eqref{result} and \eqref{dNs_prev} clearly show that both the squeezing of the input probe field and the anti-squeezing of the output one suppress the influence of the SPM, improving the sensitivity. The best possible sensitivity corresponds to the mean number of the probe photons equal to
\begin{equation}\label{N_p_opt}
  N_p = \frac{e^{R-r} + \epsilon^2e^{r-R}}{2\Gamma_S\epsilon} \,.
\end{equation}
In this case,
\begin{equation}\label{result_opt}
  (\Delta N_s)_{\rm meas}^2 = \frac{\Gamma_S}{\Gamma_X^2}\,\epsilon e^{-r-R} \,,
\end{equation}
compare with Eq.\,\eqref{dNs_min_prev}.

\section{Sensitivity estimates}\label{sec:estimates}

For the numerical estimates, we use mostly the same parameters as in Ref.\,\cite{21a1BaMaKhStIlSaLeBi}. We consider a microresonator made of ${\rm CaF}_2$, which has one of the highest intrinsic $Q$-factors $Q_{\rm intr} \simeq 3\times10^{11}$\cite{Savchenkov_OE_15_6768_2007}. In order to satisfy the requirement \eqref{Q_load}, we assume that
\begin{equation}
  Q_{\rm load} = 10^{10} \,.
\end{equation}
We assume also that the optical wavelength in vacuum is $\lambda\approx1.55\mu{\rm m}$ and the microresonator diameter is $\approx100\mu{\rm m}$. In this case, the factors $\Gamma_X$, $\Gamma_S$ can be estimated as
\begin{equation}\label{chi2G}
 	\Gamma_X = 2\Gamma_S
   = 2Q_{\rm load}\frac{n_2}{n_0} \frac{\hbar \omega_0 c}{{V_{\rm eff}}}
   \approx 0.85\times10^{-5} \,,
\end{equation}
where $c$ is the speed of light, $\omega_0$ is the optical frequency, $n_0$ is the refractive index of the material, $n_2$ is the cubic nonlinearity coefficient, $V_{\rm eff}$ is the effective volume of the mode, and $Q_{\rm load}=\omega_0\tau$ is the loaded quality factor.

For the output losses, we use the moderately optimistic value $\eta\approx0.9$.

Considering the squeezing, it should be taken into account that during the last decade, significant progress in this area was achieved, stimulated in part by the needs of the laser gravitational-wave detectors. The values of squeezing exceeding 10\,dB ($e^{-2r}=0.1$) were demonstrated, see {\it e.g.} Refs.\,\cite{Vahlbruch_PRL_117_110801_2016, Mehmet_CQG_36_015014_2019}. Therefore, we use 10\,dB for our estimates here.

It is worth noting that these values were limited mostly by the optical losses in the DOPAs. In the case of the {\it anti}-squeezing, the results are affected by the losses in much lesser degree. Therefore, here we explore higher values of the anti-squeezing factor, up to 40\,dB ($e^{2R}=10^4$). Note that the values of the optical parametric gain exceeding 40\,dB were already demonstrated experimentally, see e.g. \cite{Ooi2017_NComm_8_13878_2017}.

\begin{figure}
  \includegraphics[width=0.45\textwidth]{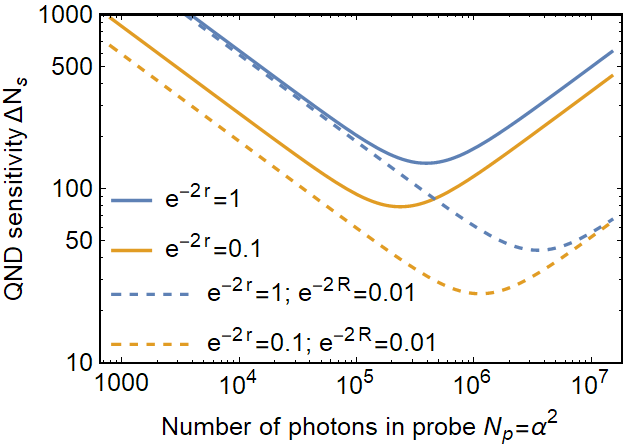}
  \caption{Plots of the measurement error $\Delta N_s$, see Eq.\,\eqref{result}, as functions of the photon number in the probe mode $N_p$ for the four characteristic combinations of the squeezing and parametric amplification factors. In all cases, $\eta=0.9$ and $\Gamma_X=2\Gamma_S=0.85\times 10^{-5}$.}\label{fig:3}
\end{figure}

In Fig.\,\ref{fig:3}, the measurement error $\Delta N_s$ is plotted as a function of the photon number in the probe mode $N_p$ for the four scenarios: (i) without squeezing and parametric amplification, (ii) with squeezing, without parametric amplification; (iii) without squeezing, with amplification; (iv) with squeezing and amplification. It can be seen from these plots that combining both these techniques, it is possible to reduce the measurement error by almost one order of magnitude, down to a few tens of photons.

\begin{figure}
  \includegraphics[width=0.45\textwidth]{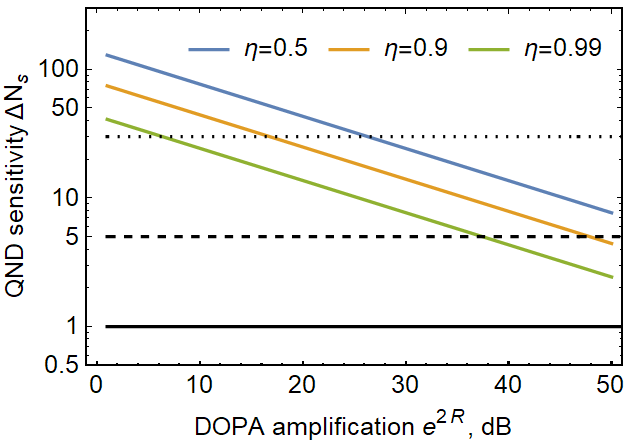}
  \caption{Plots of the optimized measurement error $\Delta N_s$, see Eq.\,\eqref{result_opt}, as functions of the parametric amplification factor $e^{2R}$. In all cases, $e^{-2r}=0.1$ (10\,dB), $\eta=0.9$, and $\Gamma_X=2\Gamma_S=0.85\times 10^{-5}$. The black solid line corresponds to the single-photon sensitivity and the black dotted line approximately corresponds to the non-Gaussianity limit.}\label{fig:5}
\end{figure}

The minima of these plots correspond to the optimal photon numbers in the probe mode, given by Eq.\,\eqref{N_p_opt}. In Fig.\,\ref{fig:5}, the corresponding optimal values of the measurement error $\Delta N_s$ are plotted as the functions of the parametric amplification factor $e^{2R}$, expressed for convenience in dB. It can be seen from these plots that the sensitivity, corresponding to preparation of the signal mode in a non-Gaussian state (see Sec.\,\ref{sec:qc}), could be achieved for the reasonably optimistic values of the parametric amplification of the probe mode.

\section{Losses in the signal mode}\label{sec:mu}

In the previous sections, we have taken into account the optical losses inside the microresonator (by mean of limiting the interaction time $\tau$) and the output losses in the probe mode. Two other mechanisms of degrading the sensitivity are the input and output losses in the signal mode. The first one is relevant for the task of measurement of the incident number of quanta, and the second one --- for preparation of the output quantum state.

\begin{figure}
  \includegraphics[width=0.4\textwidth]{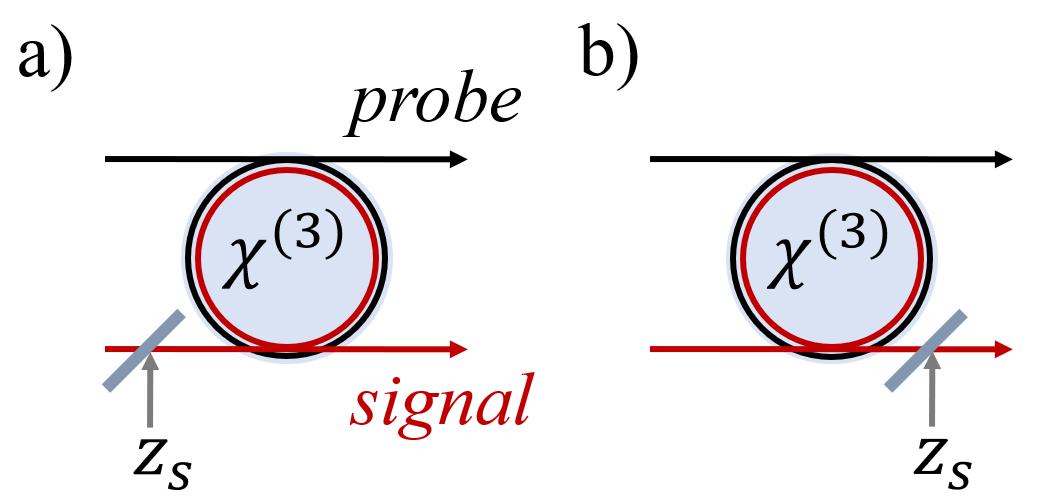}
  \caption{Accounting for losses in the signal mode: a) measurement of the incident number of quanta; b) preparation of the output quantum state. In both cases, $\mu$ is the quantum efficiency factor for the signal mode.}\label{fig:4}
\end{figure}

These two scenarios are shown in Fig.\,\ref{fig:4}. In both cases, we model the losses by means of an imaginary beamsplitter with the power transmissivity $\mu$, located, respectively, either in the input or in the output path of the signal beam.

Let us start with the first case, see Fig.\,\ref{fig:4}\,a). In this case, the imaginary beamsplitter transforms the input field as follows:
\begin{equation}
  \hat{a}_s = \sqrt{\mu}\,\hat{a}_{s\,{\rm in}} + \sqrt{1-\mu} \, \hat z_s,
\end{equation}
where the annihilation operator $\hat{a}_{s\,in }$ describes the incident field before the losses and $\hat z_s$ corresponds to a ground state field. Using this equation, it is easy to obtain the following relations for the mean numbers and the variances of the incident photon number $\hat{N}_{s\,{\rm in}} = \hat{a}_{s\,{\rm in}}^\dag\hat{a}_{s\,{\rm in}}$ and the intracavity one $\hat{N}_s=\hat{a}_s^\dag\hat{a}_s$:
\begin{subequations}\label{N_smu} \,,
  \begin{gather}
    N_s = \mu N_{s\,{\rm in}} \,, \label{N_s_loss} \\
    (\Delta N_s)^2 = \mu^2(\Delta N_s)_{\rm in}^2 + \mu(1-\mu) N_{s\,{\rm in}}
    \label{Delta_N_smu} \,,
\end{gather}
\end{subequations}
The factor $\mu$ in Eq.\,\eqref{N_s_loss} decreases the gain factor \eqref{G}: $G_{\rm loss}=\mu G$. The second term in Eq.\,\eqref{Delta_N_smu} creates an additional uncertainty in the intracavity photon number and therefore has to be added to the measurement error \eqref{result}. As a result, we obtain the following equation for the modified measurement error:
\begin{equation}\label{error}
  (\Delta N_s)^2_{\rm meas\,loss}
  = \frac{1}{\mu^2}\bigl[\mu(1-\mu)N_s + (\Delta {N_s})^2_{\rm meas}\bigr] .
\end{equation}
It is instructive to normalize it by the initial mean number of photons:
\begin{equation}
  \frac{(\Delta N_s)^2_{\rm meas\,loss}}{N_s}
  = \frac{1-\mu}{\mu} + \frac{(\Delta N_s)_{\rm meas}^2}{\mu^2N_s} .
\end{equation}
It is easy to see that the necessary condition for overcoming the SQL \eqref{SQL} is the inequality $\mu>1/2$, which corresponds to the well-known limitation on losses in optical schemes using non-classical states of light \cite{Demkowicz_PRA_88_041802_2013}.

Consider now the second task, namely the quantum state preparation, see Fig.\,\ref{fig:4}\,b). In this case, the mean value and the variance of the output photon number are equal to
\begin{subequations}\label{N_smu_prep}
  \begin{gather}
    N_{s\,{\rm prep}} = \mu N_s \,, \label{mean_N_prep} \\
    (\Delta N_s)_{\rm prep}^2 = \mu^2(\Delta N_s)_{\rm meas}^2 + \mu(1-\mu) N_s \,,
      \label{Delta_N_prep}
  \end{gather}
\end{subequations}
compare with Eqs.\,\eqref{N_smu}. We assumed here that the uncertainty of the intracavity number of quanta is equal to the measurement error $(\Delta N_s)_{\rm meas}$. Normalizing this value by the mean number of the output photons \eqref{mean_N_prep}, we obtain:
\begin{equation}
  \frac{(\Delta N_{s})^2_{\rm prep}}{N_{s\,{\rm prep}}}
  = 1-\mu\left[1-\frac{(\Delta N_s)_{\rm meas}^2}{N_s}\right] .
\end{equation}
It is easy to see that in the case of the sub-SQL intracavity sensitivity, $(\Delta N_s)_{\rm meas}<\sqrt{N_s}$, the prepared quantum state also will be a sub-Poissonian one, $(\Delta N_s)_{\rm prep} < \sqrt{N_{s\,{\rm out}}}$, for any value of $\mu$.

\section{Applications to quantum computing}\label{sec:qc}

We have shown that the modern optical microresonators allow to overcome the SQL \eqref{SQL}. Another more demanding threshold is important for the continuous-variable quantum information processing  applications \cite{Braunstein_RMP_77_513_2005}. It corresponds to the performance allowing to generate and verify quantum states characterized by non-Gaussian negative-valued Wigner quasi-probability distributions (the non-Gaussian quantum states). Note that the Gaussian states can not be orthogonal to each other \cite{Schleich2001}, while the orthogonality is necessary for many quantum phenomena. In particular, the non-Gaussian states are required for quantum computation protocols that cannot be efficiently simulated by classical computers \cite{Mari_PRL_109_230503_2012}.

It is known \cite{Bondurant_PRD_30_2548_1984, Kitagawa_PRA_34_3974_1986} that the pure quantum states with the photon number uncertainty satisfying the following inequality (in this Section, we omit for brevity all numerical factors of order of unity):
\begin{equation}
  \Delta N \lesssim N^{1/3} \,,
\end{equation}
where $N$ is the mean photon number, are non-Gaussian ones.  With account for the losses in the signal mode, see Eqs.\,\eqref{Delta_N_smu}, \eqref{Delta_N_prep}, and assuming that $1-\mu\ll1$, this inequality leads to the following requirement:
\begin{equation}
  (\Delta N_s)_{\rm meas}^2 \lesssim N_s^{2/3} - (1-\mu)N_s \,.
\end{equation}
The maximum of the R.H.S. of this condition is achieved at
\begin{equation}
  N_s \sim \frac{1}{(1-\mu)^3} \,,
\end{equation}
giving the following requirement:
\begin{equation}
  (\Delta N_s)_{\rm meas} \lesssim \frac{1}{1-\mu} \,.
\end{equation}
For the reasonably optimistic values of $\mu$, this corresponds to the measurement imprecision of a few tens of photons. According to our estimates in Sec.\,\ref{sec:estimates} (see Fig.\,\ref{fig:4}), this value can be considered as a feasible one.

The next important threshold is the single-photon QND sensitivity, which will allow to prepare and detect without absorption Fock states with arbitrary number of quanta. It follows from Eqs.\,\eqref{Delta_N_smu}, \eqref{Delta_N_prep} that in the case of $(\Delta N_s)_{\rm meas}\lesssim1$ it can be reached with values of $N_s\sim10$. It is interesting that the former requirement could be fulfilled for any value of the quantum inefficiency $\epsilon$, provided sufficiently strong squeezing (the factor $e^{-r}$) and parametric amplification (the factor $e^{-R}$). Therefore, taking into account the contemporary rapid progress in both the fabrication of high-$Q$ microresonators and the optical squeezing, it is possible to hope that the single-photon sensitivity will be implemented in a predictable future.

\section{Conclusion}\label{sec:conclusion}

We theoretically analyzed application of squeezed states of light to the considered in Ref.\,\cite{21a1BaMaKhStIlSaLeBi} scheme of quantum non-demolition measurement of optical energy, based on the effect of cross-phase modulation (XPM) in a microresonator. We showed that the sensitivity of this scheme can be radically improved using the squeezed quantum state of the probe beam and the anti-squeezing (parametric amplification) of this beam at the output of the microresonator.

We considered the sensitivity limitations imposed by optical losses in both probe and signal modes and found the optimal values of the both squeeze angles minimizing the interfering effect of self-phase modulation. We showed that for reasonably optimistic values of the optical losses in the scheme, the squeezing allows to improve the sensitivity of the QND measurement by about one order of magnitude, from a few hundreds of photons to a few tens.

Our estimates show that this sensitivity allows to generate and verify non-Gaussian quantum states, characterized by negative-valued Wigner quasi-probability distributions. Therefore, the QND measurement scheme considered here could be interesting for the quantum information processing tasks.

\section{Acknowledgment}

This work was supported by the Russian Science Foundation (project 20-12-00344). The work of S.N.B. was also supported by the Foundation for the Advancement of Theoretical Physics and Mathematics “BASIS”. The research of Sec.\,\ref{sec:qc} was supported by Rosatom in the framework of the Roadmap for Quantum computing (Contract No. 868-1.3-15/15-2021 dated October 5, 2021 and Contract No.P2154 dated November 24, 2021).

\appendix

\section{Optimization of the squeeze and homodyne angles}\label{app:angles}

Note that
\begin{equation}
  {\bf H}^{\sf T}\mathbb{S}(R,\phi) = (C\ S) \,,
\end{equation}
where
\begin{subequations}
  \begin{gather}
    C = e^R\cos(\zeta-\phi)\cos\phi - e^{-R}\sin(\zeta-\phi)\sin\phi \,, \\
    S = e^R\cos(\zeta-\phi)\sin\phi + e^{-R}\sin(\zeta-\phi)\cos\phi \,.
  \end{gather}
\end{subequations}
Therefore, it follows from Eqs.\,\eqref{G} and \eqref{d0zeta} that
\begin{gather}
  G = \sqrt{2\eta N_p}\Gamma_XS \,, \\
  \hat{d}_0^\zeta = \sqrt{\eta}(B\hat{a}_p^c + A\hat{a}_p^s) + \sqrt{1-\eta}\,\hat{y} \,,
\end{gather}
where
\begin{equation}
  A = S \,,\quad B = C + 2N_p\Gamma_SS \,.
\end{equation}

The variances of the ground fields quadratures $z^{c,s}$, $\hat{y}$ are equal to $1/2$. Therefore, with account for Eq.\,\eqref{z2a}, the variances of the quadratures $a_p^{c,s}$ and their covariance are equal to
\begin{subequations}\label{Delta_a}
  \begin{gather}
    \mean{(\Delta\hat{a}_p^c)^2} = \frac{1}{2}(\cosh2r + \sinh2r\cos2\theta)\,, \\
    \mean{(\Delta\hat{a}_p^s)^2} = \frac{1}{2}(\cosh2r - \sinh2r\cos2\theta)\,, \\
    \mean{\hat{b}_p^c\circ\hat{b}_p^s} = \frac{1}{2}\sinh2r\sin2\theta \,.
  \end{gather}
\end{subequations}
where ``$\circ$'' denotes the symmetrized product and the variance of $\hat{d}_p^\zeta$ is equal to
\begin{multline}\label{Delta_d_raw}
  \mean{(\Delta\hat{d}_0^\zeta)^2} = \frac{\eta}{2}\bigl\{
      (A^2+B^2)\cosh 2 r \\
      + [(-A^2+B^2)\cos2\theta + 2AB\sin2\theta]\sinh2r + \epsilon^2
    \bigr\} .
\end{multline}
The minimum of this equation in $\theta$ is provided by
\begin{equation}\label{theta_opt}
  \cos2\theta = \frac{A^2-B^2}{A^2+B^2}, \quad \sin2\theta =-\frac{2 A B}{A^2+B^2} \,.
\end{equation}
Taking into account that $G$ does not depend on $\theta$, it corresponds also to the minimum of \eqref{DNs}, equal to
\begin{equation}
  (\Delta N_s)^2 = \frac{(A^2+B^2)e^{-2 r} + \epsilon^2}{4N_p\Gamma_X^2S^2} .
\end{equation}

The next step is the optimization of this equation in $\zeta$. In order to simplify the equations, we use the following approach. We assume first that the second squeezing is strong enough to allow the following approximation:
\begin{equation}
  \cosh2R \approx \sinh2R \approx \frac{e^R}{2} \,.
\end{equation}
Note that this approximation holds very well even for moderate values of $R$. In this case,
\begin{equation}
  C \approx e^R\cos(\zeta-\phi)\cos\phi \,,\quad S \approx e^R\cos(\zeta-\phi)\sin\phi \,,
\end{equation}
and
\begin{multline}
  (\Delta N_s)^2 = \frac{1}{4N_p\Gamma_X^2\sin^2\phi}\biggl[
      (1 + 2N_p\Gamma_S\sin2\phi \\
      + 4N_p^2\Gamma_S^2\sin^2\phi)e^{-2r} + \frac{\epsilon^2e^{-2R}}{\cos^2(\zeta-\phi)}
    \biggr] .
\end{multline}
Evidently, the minimum of this equation is provided by
\begin{equation}
  \zeta = \phi \,.
\end{equation}
Note that for this value of $\zeta$,
\begin{equation}
  C = e^R\cos\phi \,,\quad S \approx e^R\sin\phi \,,
\end{equation}
exactly, which gives that the following equation corresponds to also exact, albeit possibly not strictly minimal value, of the measurement error:
\begin{multline}
  (\Delta N_s)^2 = \frac{1}{4N_p\Gamma_X^2}\bigl[
      (1 + \cot^2\phi + 4N_p\Gamma_S\cot\phi + 4N_p^2\Gamma_S^2)e^{-2r} \\
      + \epsilon^2(1 + \cot^2\phi)e^{-2R}
    \bigr] .
\end{multline}

The final step is the optimization in $\phi$, which is straightforward:
\begin{equation}
  \cot\phi = -\frac{2N_p\Gamma_S}{1 + \epsilon^2e^{2r-2R}}
\end{equation}
As a result, we obtain Eq.\,\eqref{result}.


%

\end{document}